\documentclass[twocolumn,showpacs,preprintnumbers,amsmath,amssymb]{revtex4}

\usepackage{graphicx}
\usepackage{dcolumn}
\usepackage{natbib}
\usepackage{bm}

\begin{document}


\title{Power-law carrier dynamics in semiconductor nanocrystals at nanosecond time scales}

\author{P. H. Sher}
\author{J. M. Smith}
\email[]{Jason.Smith@materials.ox.ac.uk} \affiliation{Department of
Materials, University of Oxford, Parks Road, Oxford OX1 3PH, UK}

\author{P. A. Dalgarno}
\author{R. J. Warburton}
\affiliation{School of Engineering and Physical Sciences,
Heriot-Watt University, Edinburgh EH14 4AS, UK}

\author{X. Chen}
\author{P. J. Dobson}
\affiliation{Department of Engineering Science, University of
Oxford, Parks Road, OX1 3PJ,UK}

\author{S. M. Daniels}
\author{N. L. Pickett}
\author{P. O'Brien}
\affiliation{School of Chemistry, University of Manchester, Oxford
Road, Manchester M13 3PL, UK}

\date{\today}

\begin{abstract}
We report the observation of power law dynamics on nanosecond to
microsecond time scales in the fluorescence decay from semiconductor
nanocrystals, and draw a comparison between this behavior and power-law fluorescence blinking from single nanocrystals. The link is supported by comparison of blinking and lifetime data measured simultaneously from the same nanocrystal. Our results reveal that the power law coefficient changes little over the nine decades in time from 10 ns to 10 s, in contrast with the predictions of some diffusion based models of power law behavior.
\end{abstract}


\maketitle

The following article has been submitted to Applied Physics Letters.  If it is published, it will be found online at http://apl.aip.org


Power law dynamics observed in the fluorescence blinking behavior of
single semiconductor nanocrystals have received much recent
attention. Both the fluorescence \textit{on} and \textit{off}
periods have been shown to follow power law distributions, with the probability density of measuring a period of duration $\tau$ given by $P(\tau)=\mu \tau_{b}^\mu \tau^{-(1+\mu)}$, where $\mu$ is the time independent power law coefficient, and $\tau_b$ is the temporal resolution of the measurement apparatus. The observed behaviour is particularly robust for the fluorescence \textit{off} durations which show values of $\mu$ between about 0.4 and 0.8 over time scales from 100~$\mu$s to 100~s \cite{Kuno00,Shimizu01,Tang05JCP}. Autocorrelation measurements of ensemble fluorescence have suggested that the behavior extends to time scales as short as 1~$\mu$s \cite{Verberk02}. The behavior is of significant practical importance, since the occurrence of long fluorescence \textit{off} durations leads to statistical ageing (reversible photobleaching) in the fluorescence from ensembles of emitters \cite{Brokmann03}, raising serious obstacles to the realization of inexpensive optoelectronic devices based on nanocrystal/polymer composites. 

The blinking itself is thought to be caused by the hopping of carriers between the quantum confined state in the nanocrystal and surrounding trap sites: when
the nanocrystal is in a charged state, fluorescence is quenched by
strong non-radiative Auger processes facilitated by the extra
confined carrier. However the origin of the power law behavior is
still a matter of some debate. Several suggestions have been made
for dynamic models that rely on a random walk or diffusion process
\cite{Shimizu01,Tang05JCP,Tang05PRL,Frantsuzov05,Margolin04}, whilst
others propose static models with a distribution of trap states in
the volume surrounding each nanocrystal \cite{Verberk02,Kuno03}. Some works \cite{Tang05PRL, Frantsuzov05} have suggested that a signature of a diffusion process would be a reduction in the power law gradient at
sub-100 $\mu$s time scales such that $-1<\mu<0$.

In this Letter we present evidence that power law behavior can be
observed on nanosecond to microsecond time scales in time-resolved
photoluminescence (TRPL) experiments performed on both single
nanocrystals and ensembles, thereby providing a new time scale on
which to study the phenomenon. We used a standard time-correlated
single photon counting (TCSPC) apparatus with timing resolution
$t_{res} \simeq 500$~ps to measure room temperature TRPL of three different nanocrystal types: bare CdSe nanocrystals
(sample A); CdSe/ZnS core-shell structures (sample B); and
$\text{Zn}_{0.5}\text{Cd}_{0.5}\text{S/CdSe/ZnS}$ triple layer structures (sample C). The core-shells were
fabricated by standard methods; the fabrication procedure for the
bare nanocrystals and the triple layer structures are described elsewhere
\cite{XCthesis,SMDthesis}. All of the samples showed strong ensemble
luminescence at around 600 nm with no deep
trap luminescence, but importantly were not optimized for high
quantum yield so that long fluorescence \textit{on} times were
absent. Samples for photoluminescence measurements were created by
spin casting from solution onto quartz substrates. Single
nanocrystals were identified by their homogeneously broadened PL
spectra with Lorentzian line shapes.

\begin{figure}
\includegraphics[width=7cm]{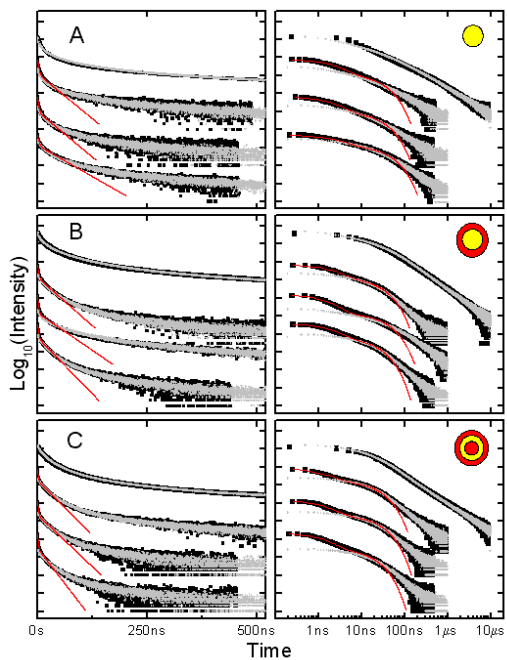}
\caption{\label{Fig. 1} (color online) TRPL data from semiconductor nanocrystals on semi-logarithmic (left column) and logarithmic (right column) axes.
The three rows correspond to the different nanocrystal structures A,
B and C. All data sets are from single nanocrystals except for the
uppermost in each window which is from an ensemble. The black
scatter plots are raw data, the gray scatter plots are results of
our Monte Carlo simulation, and the solid red lines are analytical
biexponential fits. In all cases, $t = 0$ corresponds to the peak of
the TRPL signal.}
\end{figure}

Figure \ref{Fig. 1} shows fluorescence decay data for three single
nanocrystals and an ensemble of each of the sample types A, B, and C. The same data (black squares) are displayed on semi-logarithmic axis in the left hand column and fully logarithmic axes in the right hand column of the figure. Power law behaviour appears as a straight line of gradient $-(\mu +1 )$ on the fully logarithmic axes. Previously, single nanocrystal fluorescence decays have been fitted with biexponential functions with lifetimes of about 1~ns and 15~ns,
attributed to Auger and radiative recombination processes respectively \cite{Schlegel02}. Fitting biexponential functions to our data (solid red lines in Fig.\ref{Fig. 1}) reveals excellent agreement up to about 50~ns after the excitation pulse, but for longer delays the fluorescence consistently decays more slowly than the biexponential function.

The presence of this slow tail indicates that between excitation and
emission, the photogenerated exciton can enter a `dark' state from
which it can not radiate. We do not believe this state to be the `dark' spin configuration of the confined exciton ground state since the lowest energy `bright' exciton states for nanocrystals of this size lie less than 5 meV
($\ll kT$) higher in energy \cite{Norris96}, and the spin relaxation
time is fast compared with the radiative lifetime
\cite{Gupta02,Crooker03}. That the tail results from a large
homogeneous line width as suggested in the recent work by Rothe et al
\cite{Rothe06} appears unlikely since homogeneous broadening at room temperature  is dominated by rapid dephasing of the exciton and not population decay. We therefore believe that the dark state occurs when one of the
carriers escapes from the nanocrystal, similarly to the fluorescence \textit{off} state that is responsible for the single nanocrystal blinking phenomenon.

\begin{figure}
\includegraphics[width=7cm]{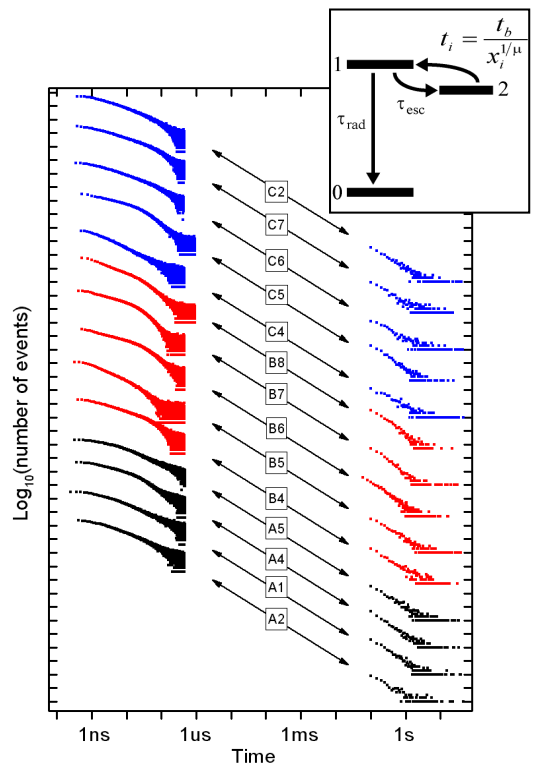}
\caption{\label{Fig. 2} (color online) TCSPC and blinking data histograms on common logarithmic axes for fourteen different single nanocrystals.
The increments on both axes of the figure correspond to factors of
ten. (Inset) three level system used to simulate the
tails of the TRPL data. Level 0 represents the nanocrystal ground
state; level 1 the quantum confined exciton; and level 2 the charged
fluorescence \textit{off} state. $\tau_{rad}$ is the radiative
lifetime that corresponds to the slower decay in the original
biexponential fit. The two new fitting parameters are the lifetime
for escape from the exciton state to the fluorescence \textit{off}
state, labeled $\tau_{esc}$, and the power law coefficient $\mu$.}
\end{figure}

An empirical Monte Carlo (MC) model that produces the suggested power-law behavior is shown schematically in the inset of Fig.\ref{Fig. 2}. Times spent in the fluorescence \textit{off} state are generated using the function $t_i=t_{b}x_i^{-1/\mu}$ where $x_i$ is a random variable in the range $0 < x_i < 1$, and $t_{b}$ is the minimum measurable trapping duration \cite{Bardou01}, which in our experiment is equal to $t_{res}$. For each MC iteration the system
is initialized in the exciton state at time $t = 0$ and is allowed to evolve until it undergoes radiative recombination to the ground state.

Fits of our MC simulation data sets to the measured TRPL data are
shown as grey scatter graphs overlaying the experimental data in
Fig.\ref{Fig. 1}. In all cases the model provides exact fits to the
experimental data for time delays greater than about 10 ns. The
underestimation of the single nanocrystal experimental data at short time delays results because our model neglects Auger limited fluorescence. The fitting parameters for the twelve MC data sets in Fig.\ref{Fig. 1} are listed in Table \ref{tab:table1}. In each case we have taken $\mu$ value from the ensemble fit. The tabulated parameters reveal that most of the differences between the data sets are primarily attributable to differences in $\tau_{esc}$, which is determined by
the degree of surface passivation of the individual nanocrystals.
That all values of $\tau_{esc}$ fitted are smaller than $\tau_{rad}$
shows that the \textit{off} state plays an important role. 

\begin{table}
\caption{\label{tab:table1} Fitting parameters $\tau_{rad}$,
$\tau_{esc}$, and $\mu$ for the best fits to the data shown in Fig.
\ref{Fig. 1}, and for the average $\mu$ values from the blinking
data in Fig. \ref{Fig. 2}. The numbers in parentheses represent
estimated error bars for the values stated.\\}
\begin{ruledtabular}
\begin{tabular}{lcccc}
Sample&$\tau_{rad}(ns)$ & $\tau_{esc}(ns)$ & $\mu$ & $\mu_{blink}$\\
 \hline
Aens & 14 (2) & 2 (1) & 0.50 (0.05) & 0.54 (0.05)\\
A1 & 14 (2) & 2.5 (1) & 0.50 (0.1)\\
A2 & 14 (2) & 4 (1) & 0.50 (0.1)\\
A3 & 20 (2) & 5 (1) & 0.50 (0.1)\\
Bens & 14 (2) & 2.5 (1) & 0.63 (0.05) & 0.67 (0.05)\\
B1 & 14 (2) & 7 (1) & 0.63 (0.1)\\
B2 & 20 (2) & 3 (1) & 0.63 (0.1)\\
B3 & 12 (2) & 7 (1) & 0.63 (0.1)\\
Cens & 17 (2) & 6 (1) & 0.50 (0.05) & 0.37 (0.05)\\
C1 & 17 (2) & 10 (2) & 0.50 (0.1)\\
C2 & 17 (2) & 10 (2) & 0.50 (0.1)\\
C3 & 17 (2) & 10 (2) & 0.50 (0.1)\\
\end{tabular}
\end{ruledtabular}
\end{table}

To probe the relationship between the fluorescence decay function
and distribution of \textit{off} times from blinking measurements,
we performed both experiments simultaneously on the same
nanocrystal. Pulsed excitation at a repetition rate of 500 kHz allowed us to
use a TRPL time range of 1 $\mu$s and a blinking sampling time of 50
ms with clear distinction between the fluorescence \textit{on} and \textit{off} states. Figure 3 shows both TCSPC and fluorescence \textit{off}
histograms for fourteen single nanocrystals of the different types.

We have scaled the TCSPC data in Fig. \ref{Fig. 2} to enable the two data sets to be presented on the same $y$-axis. The scaling
factor used is the product of two sub-factors --the first, $F_1$,
corrects for the difference between the widths of the time bins,
generally 50 ms for blinking and 200 ps for TCSPC, giving $F_1 = 2.5\times10^8$, while the second, $F_2$, corrects for the difference in the
measurement efficiency of the two methods. Whilst the
blinking data records all of the \textit{off} periods with duration
$>$ 100 ms, the TCSPC data records a timing event for only about 4$\%$ of the excitation pulses, so that $F_2 \simeq$ 25. The result of this
scaling is therefore to shift the TCSPC data up the $y$-axis of the
logarithmic plot by about ten decades without distortion to the curve shape.

Although there is clearly a large time gap between the two data sets
and some variation between individual nanocrystals, in each case
their relative positions suggest that the power law decay at the
short and long time scales measured here may be continuous across
the entire time range with $\mu \simeq$ 0.5 throughout. There is
also some correlation between the values of $\mu$
measured on the different time scales. The mean values measured from
the blinking data, listed alongside the ensemble fitting parameters
for each of the three nanocrystal types in Table \ref{tab:table1},
agree well with those fitted to the ensemble TRPL tails, with
nanocrystals of type B (core-shells) providing a noticeably larger
value of $\mu$ in both measurements. This degree of correspondence
between the two data sets provides further evidence that the tail in
the fluorescence decay is a result of the same trapping behavior
that causes fluorescence blinking. We note that in contrast
with the predictions made in the diffusion based models of
references \cite{Tang05PRL} and \cite{Frantsuzov05} the values of
$\mu$ appear quite similar on the 100 ns time scale to those found
in the blinking data.

To conclude, our study provides evidence that the same carrier
trapping dynamics that lead to power law statistics in fluorescence
blinking of single semiconductor nanocrystals can be observed on
much faster time scales in fluorescence decay experiments, opening
up new possibilities for testing the various theoretical models of
power law behavior. We hope that as well furthering the
understanding of carrier dynamics in semiconductor nanocrystals, our
approach may prove fruitful in the analysis of the non-exponential
fluorescence decays of other
colloidal and molecular systems.

\begin{acknowledgments}
Funding for this work was provided by the United Kingdom Engineering
and Physical Sciences Research Council, the Royal Society of
Edinburgh, the Royal Society of London, and Oxford University's
Research Development Fund. We would like to thank PicoQuant GmbH for
the loan of a pulsed laser for the single nanocrystal measurements.
\end{acknowledgments}


\end{document}